\documentclass[pdflatex,sn-mathphys-num]{sn-jnl}

\usepackage{graphicx}%
\usepackage{multirow}%
\usepackage{amsmath,amssymb,amsfonts}%
\usepackage{amsthm}%
\usepackage{mathrsfs}%
\usepackage[title]{appendix}%
\usepackage{xcolor}%
\usepackage{textcomp}%
\usepackage{manyfoot}%
\usepackage{booktabs}%
\usepackage{algorithm}%
\usepackage{algorithmicx}%
\usepackage{algpseudocode}%
\usepackage{listings}%
\usepackage{booktabs}%
\usepackage{tikz}
\usepackage{amsmath}
\usepackage{overpic}
\newcommand{\lsline}[1]{%
  \tikz[baseline=-0.6ex]\draw[line width=0.7pt,#1] (0,0) -- (2em,0);}

\DeclareRobustCommand{\lsSolid}{\lsline{}}
\DeclareRobustCommand{\lsDotted}{\lsline{dotted}}
\DeclareRobustCommand{\lsSparse}{\lsline{dash pattern=on 0.7pt off 3pt}}
\DeclareRobustCommand{\lsDashed}{\lsline{dashed}}
\DeclareRobustCommand{\lsLong}{\lsline{dash pattern=on 6pt off 3pt}}
\DeclareRobustCommand{\lsDashDot}{\lsline{dash dot}}
\DeclareRobustCommand{\lsDashDotDot}{\lsline{dash dot dot}}                                                    

\theoremstyle{thmstyleone}%
\theoremstyle{thmstyletwo}%

\theoremstyle{thmstylethree}%

\begin{document}

\title[Article Title]{Effects of Prandtl number variation on turbulent forced convection over porous lattices}


\author*[1]{\fnm{Aneek} \sur{Chakraborty}}\email{a.chakraborty@tudelft.nl}

\author[1]{\fnm{Stefan} \sur{Hickel}}\email{s.hickel@tudelft.nl}
\equalcont{These authors contributed equally to this work.}

\author[2]{\fnm{Davide} \sur{Modesti}}\email{davide.modesti@gssi.it}
\equalcont{These authors contributed equally to this work.}

\affil*[1]{\orgdiv{Aerodynamics Group, Faculty of Aerospace Engineering}, \orgname{Delft University of Technology}, \orgaddress{\street{ Kluyverweg 2}, \city{Delft}, \postcode{2629HS}, \country{ The Netherlands}}}

\affil[2]{ \orgname{Gran Sasso Science Institute}, \orgaddress{\street{viale Francesco Crispi 7}, \city{L'Aquila}, \postcode{67100}, \country{Italy}}}


\abstract
{Recent studies on heat transfer at rough walls~(Rowin et al.,  Journal of Fluid Mechanics, 2024) show that there are distinct contributions from exposed and sheltered regions, each having a different Prandtl-number scaling. We test whether this holds also for porous surfaces using pore-resolved direct numerical simulations of turbulent channel flow over cubic-lattice substrates with porosities of $71\%$ and $87\%$, Prandtl numbers ranging from $Pr = 0.01$ to $4$, and friction Reynolds number $Re_\tau \approx 700$. At $71\%$ porosity and $Pr > 0.1$, the exposed and sheltered coefficients scale as $Pr^{-0.73}$ and $Pr^{-0.52}$, close to rough-wall laws. At the higher porosity, $\Phi=87\%$, turbulence penetrating the pores affects the sheltering around the ligaments, and all coefficients scale with $Pr^{-2/3}$. For $Pr < 0.1$, conduction dominates and all coefficients scale as $Pr^{-1}$. We further propagate the local heat transfer coefficient to predict the bulk Stanton number and discuss the limitations of this approach.}


\maketitle

\section{Introduction}\label{introduction}
Turbulent forced convection in internal flows is central to several industrial applications, such as fuel cells, nuclear reactors, turbine blades and rocket nozzles. In these applications, heat exchangers are a key component, and their primary purpose is to maximize heat transfer. Surface texturing intensifies turbulent mixing and has therefore become a common strategy in industry for achieving this goal.

It is well established that surface undulations, in the form of either roughness or porous substrates, increase not only heat transfer but, in most cases, also drag~\citep{Owen_Thomson_1963, Yaglom_Kader_1974, Brutsaert1975, MacDonald_Hutchins_Chung_2019, Chungreview, Chakraborty2026}. The latter is generally undesirable in industrial applications as it is directly related to an increase in required pumping power.  The heat-transfer increase can be quantified through the viscous-scaled downward shift of the mean temperature profile relative to the smooth wall, expressed by the temperature deficit $\Delta \Theta^+=\Theta_S^+-\Theta_R^+$. Subscripts S and R refer to the smooth and rough/porous walls, respectively.
 The plus superscript denotes quantities scaled in wall units, using the friction velocity $u_\tau=\sqrt{\tau_w/\rho}$, the viscous length scale $\delta_v=\nu/u_\tau$ and the friction temperature $\theta_\tau=q_w/(\rho c_p u_\tau)$, where $\rho$ is the fluid density, $\tau_w$ is the wall shear stress and $q_w$ is the wall heat flux. The temperature deficit, $\Delta \Theta^+$, is the thermal counterpart of the Hama roughness function $\Delta U^+=U_S^+-U_R^+$ for momentum, such that
\begin{subequations}
\begin{align}
U_R^+ &= \frac{1}{\kappa}\log{(y^+ + \ell_T^+)} + A - \Delta U^+ (\ell^+), \label{eq:vel_R}\\
\Theta_R^+ &= \frac{1}{\kappa_\theta}\log{(y^+ + \ell_{\theta}^+)} + A_\theta(Pr) - \Delta \Theta^+ (\ell^+,Pr), \label{eq:temp_R}
\end{align}
\end{subequations}
where $U_R^+$ and $\Theta_R^+$ are the rough-wall mean velocity and temperature, and $\kappa=0.387$ and $A=4.8$ are the von K\'arm\'an constant and the additive constant of the mean velocity profile in channel flow. The roughness lengthscale is $\ell^+$, and $\ell_T^+$ and $\ell_{\theta}^+$ are the virtual origin shifts for momentum and temperature, respectively.    For the temperature, $\kappa_\theta=0.459$, and the smooth-wall intercept $A_\theta$ depends on the Prandtl number $Pr=\nu/\alpha$,
\begin{equation}\label{calcBeta}
     A_{\theta}(Pr) = \frac{1}{\kappa_{\theta}} \left [\frac{2\pi\ C_{\theta}^{2/3}}{3\sqrt{3}} Pr^{2/3} + \frac{1}{3} \log Pr - \left(\frac{1}{6} +\frac{1}{2\sqrt{3}} +\frac{2}{3} \log C_{\theta} -\log \kappa_{\theta}  \right)\right],
\end{equation}
where $C_{\theta} = 10$~\citep{pirozzoli_23,Pirozolli_Modesti_2024}, and $\nu$ and $\alpha$ are the kinematic viscosity and thermal diffusivity of the fluid. Here, $\Theta=T-T_w$ is the temperature difference with respect to the wall temperature $T_w$. We use capital letters to denote Reynolds averages in time and in the homogeneous directions, and lowercase letters for the corresponding fluctuations. 

Understanding the mechanism behind the heat-transfer increase over rough and porous walls requires accurate local measurements. Temperature measurements, however, are inherently challenging and remain scarce~\citep{GOWEN19671701, SUBRAMANIAN19811641}, and the difficulty increases further over rough and porous surfaces. Experiments over rough walls have shown that resolving near-wall thermal structures demands careful measurement techniques, particularly when temperature fluctuations must be captured~\citep{AbuRowin2024,Foroozan2025}. Porous surfaces add further complexity, as fluid penetration into the substrate gives rise to sub-surface transport that is hard to access with conventional wall-based techniques.

Direct numerical simulation (DNS) therefore offers an attractive route to studying turbulent forced convection over rough and porous surfaces. DNS has been used extensively to examine how turbulent flow behaves over rough walls~\citep{Raupach1991, Jimenez2004, MacDonald_Hutchins_Chung_2019, PEETERS2019454, Zhong_Hutchins_Chung_2023} and porous walls~\citep{Breugem, breugem_boersma_uittenbogaard_2006, Chandesris, NishiyamaKuwataSuga, Brethouwer_Habibi_Khorasani_Bagheri_2026}. In contrast, few studies have examined the thermal field over porous surfaces, and those available provide little insight into how the porous substrate affects heat transfer locally. Although there is consensus that rough and porous surfaces increase drag and heat transfer, as quantified by $\Delta U^+$ and $\Delta \Theta^+$, respectively, the understanding of the mechanism underlying the heat-transfer increase has advanced considerably less than for momentum. At a rough or porous surface, heat transfer is dominated by molecular conduction and is governed by the thickness of the thermal boundary layer $\delta_{\theta}$, which varies with the roughness type, the permeability of the porous surface and $Pr$. The temperature deficit alone is therefore not sufficient to determine how the surface topology or the fluid properties influence heat transfer locally.

\citet{Abu_Rowin_Zhong_Saurav_Jelly_Hutchins_Chung_2024} examined local heat transfer over rough walls using a systematic set of DNS of three-dimensional sinusoidal roughness with varying frontal solidity, for $Pr=0.5$-$2$. They showed that the analogy between heat and momentum transfer breaks down as the fully rough regime is approached. To explain this behaviour, they partitioned the rough surface into two types of region: exposed, high-shear regions, where heat transfer is enhanced and follows a local Reynolds analogy, and sheltered regions of reversed flow, where heat transfer is weaker and nearly uniform in space. The exposed heat-transfer coefficient $C_{h,e}$ follows the classical $Pr^{-2/3}$ scaling of the Reynolds analogy, whereas the sheltered coefficient $C_{h,s}$ scales as $Pr^{-0.45}$. The overall heat transfer is then the sum of the two contributions, weighted by the exposed and sheltered wetted areas. Building on this exposed-sheltered decomposition, the authors formulated a model for the rough-wall heat-transfer coefficient as a function of the roughness Reynolds number, the solidity and $Pr$. However, they considered only $Pr \sim \mathcal{O}(1)$, for which heat transfer is primarily driven by advection. The behaviour of the local heat transfer and how it scales at low $Pr$ ($Pr \sim \mathcal{O}(10^{-2})$), where the thermal field is conduction-dominated, remains an open question.

In a recent study~\citep{Chakraborty2026}, we performed DNS of turbulent channel flow over cubic-lattice porous substrates, for a wide range of $Re_{\tau}$, volume porosities and $ Pr$, with temperature treated as a passive scalar. We developed predictive models for the friction coefficient $C_f$ and the Stanton number $St$ that agree with the DNS data to within $5\%$. That study, however, did not address how heat transfer behaves locally within the porous substrate or at the porous-fluid interface. Extracting this information requires fully resolved velocity and temperature fields, for which the time-averaged three-dimensional fields from DNS are well suited.

In the present work, we use time-averaged flow fields from pore-resolved DNS to investigate the local heat-transfer characteristics within porous substrates over a wide range of $Pr$. We assess whether the scaling laws derived for local heat-transfer coefficients over rough walls also hold for porous surfaces, and we propagate the local heat-transfer coefficient to a prediction of the bulk Stanton number $St$.

\section{Methodology}\label{methodology}

\begin{figure}
    \centering
    \begin{overpic}[width=\textwidth]{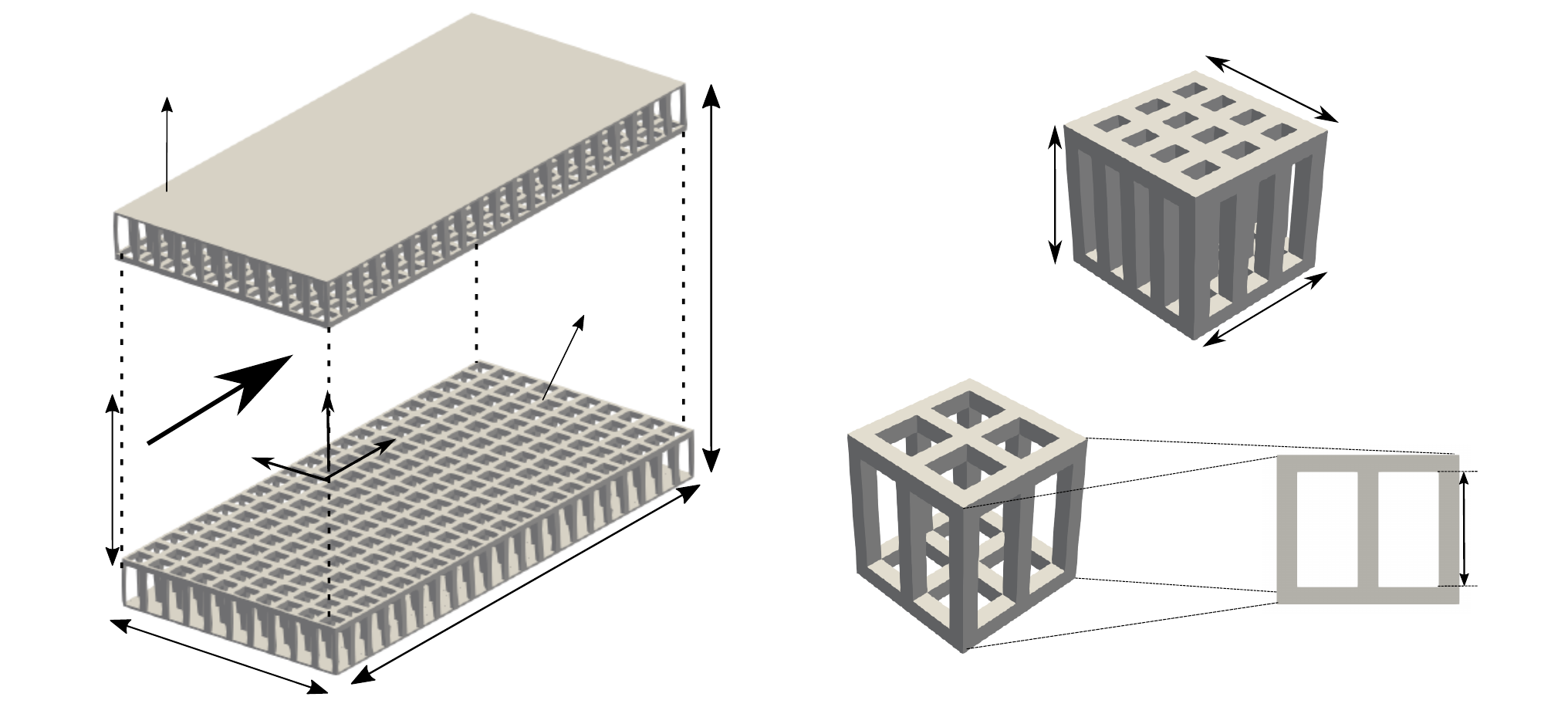}
     \put(2,45){$(a)$}
     \put(12,1){\small $L_z$}
        \put(33,5.5){\small $L_x$}
        \put(46,26){\small $L_y$}
        \put(8,21){\textbf{Flow}}
        \put(33,28.5){\small porous}
        \put(33,26.5){\small substrate}
        \put(4,13){ $\delta$}
        \put(4,6){ \small $h$}
        \put(26,16) {$\mathbf{x}$}
        \put(19,20) {$\mathbf{y}$}
        \put(13,14.5) {$\mathbf{z}$}
        \put(8.5,39){\small wall}
     \put(46,45){$(b)$}
        
        \put(62,40){\small (i)}
        \put(50,20){\small (ii)}
        \put(62,32){\small $0.3\delta$}
        \put(81,40){\small $0.3\delta$}
        \put(81,23){\small $0.3\delta$}
        \put(94,11){\small $d_i$}
    \end{overpic}
    \caption{\textit{(a)} Schematic of the computational domain with dimensions $L_x\times L_y\times L_z = 3\delta\times 2.6\delta\times 1.5\delta$. \textit{(b)} Unit cells of porous geometries used for simulations for (i) $71\%$ and (ii) $87\%$. The unit cell has dimension $0.3\delta\times 0.3\delta\times 0.3\delta$ and $d_i$ is the pore size in the i-th direction.}
    \label{fig:SCCgeo}
\end{figure}

\begin{table}[htbp]
\centering
\caption{Summary of the flow case parameters for the current study. Cases with prefix S refer to the smooth wall. The viscous scaled streamwise ($\Delta x^+$) and spanwise ($\Delta z^+$) grid spacing, wall-normal grid spacing at the porous interface ($\Delta y^+_\mathrm{wall}$), the equivalent sandgrain roughness for the cubic lattice geometry $k_s^+ = 4.35/\alpha_x^+$~\citep{Chakraborty2026}, the temperature deficit $\Delta\Theta^+$, the Stanton number $St$ and the Nusselt number $Nu$ are reported for each case. }
\label{tab:results}
\begin{tabular}{lccccccccc}
\toprule
Case & $Pr$ & $Re_\tau$ &$\Delta x^+$& $\Delta z^+$& $\Delta y^+_\mathrm{wall}$ &$k_s^+ = 4.35/\alpha_x^+$& $\Delta\Theta^+$ & $St(\times 10^3)$ & $Nu$ \\
\midrule
SL1 & 0.01 & 677 &5.28&2.64&0.69 &-& - & 25.05 & 6.51 \\
SL2 & 0.05 & 677 &5.28&2.64&0.69  &-& - & 9.35 & 12.16 \\
SL3 & 0.1 & 677 &5.28&2.64 &0.69&-& - & 7.09 & 18.45 \\
SM1 & 0.5 & 677 &5.28&2.64 &0.69&-& - & 3.74 & 48.69 \\
SM2 & 1.0 & 677 &5.28&2.64 &0.69&-& - & 2.72 & 70.67 \\
SM3 & 2.0 & 677 &5.28&2.64 &0.69&-& - & 1.92 & 99.83 \\
SH1 & 4.0 & 683 &2.67 &1.33 &0.35&-& - & 1.37 & 142.07 \\
L1 & 0.01 & 689  &2.01&2.06  &0.77&30.67& 0.05 & 37.49 & 7.89 \\
L2 & 0.05 & 689  &2.01&2.06 &0.77&30.67& 0.23 & 14.29 & 15.05 \\
L3 & 0.1 & 689  &2.01&2.06 &0.77&30.67& 0.44 & 11.09 & 23.36 \\
M1 & 0.5 & 691  &2.02&2.06&0.77&30.67& 2.37 & 6.71 & 70.69 \\
M2 & 1.0 & 691  &2.02&2.06&0.77&30.67& 5.02 & 5.49 & 115.79 \\
M3 & 2.0 & 691  &2.02&2.06&0.77&30.67& 10.08 & 4.58 & 193.03 \\
H1 & 4.0 & 688  &1.34&1.37&0.64&30.67& 18.69 & 4.02 & 338.83 \\
M1-87 & 0.5 & 694 &2.13&2.03 &0.78&173.74& 5.95 & 16.77 & 117.28 \\
M2-87 & 1.0 & 694 &2.13&2.03 &0.78&173.74& 10.11 & 14.37 & 201.00 \\
M3-87 & 2.0 & 694 &2.13&2.03 &0.78&173.74& 15.86 & 11.39 & 318.68 \\
\bottomrule
\end{tabular}
\end{table}

We solve the incompressible Navier-Stokes equations, given by:
\begin{gather}
    \nabla \cdot \mathbf{u} = 0,\\
    \frac{\partial{\mathbf{u}}}{\partial{t}} + \nabla\cdot\mathbf{uu} = -\nabla p +\frac{1}{Re}\nabla^2\mathbf{u} + \Pi,
\end{gather}
and the passive scalar equation for temperature:
\begin{gather}
     \frac{\partial \theta}{\partial t} + \nabla\cdot\mathbf{u}\theta=\frac{1}{Re \cdot Pr}\nabla^2 \theta + Q,
\end{gather}
where $\mathbf{u}$ is the velocity vector, $p$ is the pressure term, $\nabla$ is the gradient operator, $Re$ is the Reynolds number, $t$ is time, $\Pi$ is the fluctuating pressure gradient, $\theta$ is non-dimensionalized temperature and $Q$ is the source term for balancing heat losses.

The equations are
discretised with a second-order finite-difference method on a staggered mesh~\citep{orlandi_00}.
The Poisson equation for pressure is solved in spectral space in the homogeneous
directions, whereas a second-order finite-difference discretisation is used in the
wall-normal direction. For time integration, the third-order, three-stage
Runge-Kutta algorithm by Alan Wray is employed. The streamwise momentum equation is
forced to maintain a constant mass flow rate. Similarly, each scalar is forced by a
uniform volumetric heat source that maintains a constant bulk temperature. Periodic
boundary conditions are applied in the streamwise and spanwise directions, whereas
no-slip, isothermal conditions are imposed on all solid surfaces, including
the ligaments of the porous medium. The ligaments are represented with the
immersed boundary method of \citet{luchini2025}, applied to both the velocity
and scalar fields.\par

Cubic-lattice structures are used as porous substrates, following
\citet{Kuwata_Suga_2017, Habibi_Khorasani_Luhar_Bagheri_2024} and \citet{ Chakraborty2026}, as shown in figure~\ref{fig:SCCgeo}. The ligaments have a square cross-section with side length
$d_l = 0.039\delta$, which is kept fixed throughout this study. The volume porosity
$\Phi$, defined as the ratio of fluid volume to total volume of the substrate, is
set by controlling the pore sizes $d_x$, $d_y$ and $d_z$ in the streamwise,
wall-normal and spanwise directions, respectively. We consider two porosities,
$\Phi = 0.71$ and $\Phi = 0.87$, at a fixed friction Reynolds number
$Re_\tau \approx 700$. For $\Phi = 0.71$, the Prandtl number range includes low
($Pr = 0.01, 0.05, 0.1$), moderate ($Pr = 0.5, 1, 2$) and high ($Pr = 4$)
values. For $\Phi = 0.87$, we simulate the moderate range $Pr = 0.5$-$2$, so
that the effect of porosity can be isolated. We also performed baseline smooth-wall
simulations at approximately matching $Re_\tau$, as reported in
table~\ref{tab:results}. 

Simulations are performed in a domain with dimensions
$3\delta \times 2(\delta+h) \times 1.5\delta$, where $\delta$ is the
half-channel height and $h$ is the height of the porous substrate. A uniform grid
spacing is used in the periodic directions, whereas the grid-stretching function
of \citet{PIROZZOLI2021110408} is used to cluster points in the wall-normal
direction near the porous-fluid interface. A geometric progression is used to
increase the mesh spacing towards the bottom of the substrate. The wall-parallel mesh spacing gives about seven grid points per
ligament and resolves the Batchelor scale of the temperature field up to
$Pr = 2$. For $Pr = 4$, the mesh is refined by a further factor of
$\sqrt{2}$ in each wall-parallel direction, because the Batchelor scale
scales as $\eta_B \propto Pr^{-1/2}$. It should be noted here that the grid cell size to Batchelor scale ratio at $Pr=4$ reaches a maximum of 1.2 in the entire domain, which shows that the thermal scales are well resolved. \par

Statistics are collected for at least
$\Delta T\,u_\tau/\delta = 50$, after the flow fields have reached a fully-developed, statistically stationary state and time-averaged
three-dimensional fields of velocity, pressure and temperature are stored for every
$Pr$. We denote time averaging by $\langle\cdot\rangle_t$ and use an overbar to denote
intrinsic Reynolds averaging, i.e. averaging in time and in the homogeneous
directions over the fluid region only,
\begin{equation}
  \overline{f}(y) = \frac{1}{A_f(y)}\int_{A_f(y)} \langle f \rangle_t \,\mathrm{d}A ,
\end{equation}
where $A_f(y)$ is the fluid area of the wall-parallel plane at height $y$.\par

\section{Results}\label{results}
\subsection{Instantaneous thermal field}

\begin{figure}
\centerline{\includegraphics{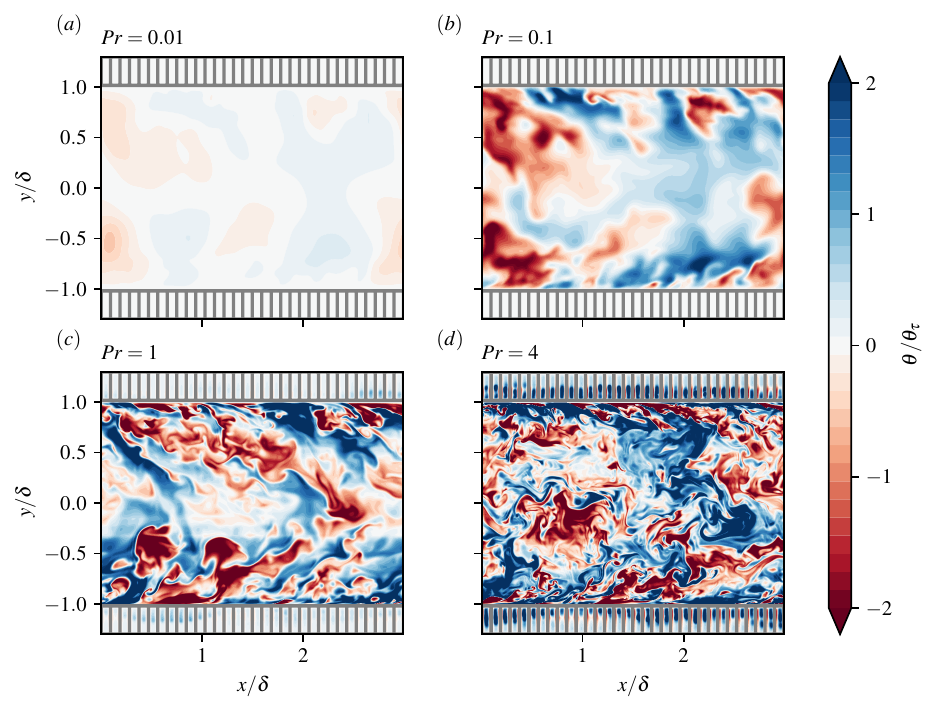}}
  \caption{Instantaneous temperature fluctuations $\theta/\theta_{\tau}$ in a longitudinal plane for cases $L1$ \textit{(a)}, $L3$ \textit{(b)}, $M2$ \textit{(c)} and $H1$ \textit{(d)}.}
\label{fig:tempfluccontours}
\end{figure}

We begin by examining the instantaneous temperature fluctuations, normalised by the friction temperature, in figure~\ref{fig:tempfluccontours}. The thermal structures change markedly with $Pr$, both in amplitude and in the range of scales. At $Pr=0.01$, the high thermal diffusivity smoothens out the finer thermal structures, and the thermal field is dominated by large, weak conductive patches that do not penetrate into the porous substrate. At $Pr=1$, the temperature field closely resembles the velocity field, with inclined thermal structures spanning half the channel height. A thermal layer also forms at the porous-fluid interface, and weak fluctuations penetrate into the porous matrix. At $Pr=4$, the thermal field is advection-dominated and is characterised by thinner structures with sharper gradients, which we attribute to the thermal diffusive length scale being smaller than the Kolmogorov scale. The near-wall structures are thinner, and the fluctuations penetrate deeper into the porous substrate, with alternating positive and negative fluctuations between the cavities. This indicates enhanced thermal exchange between the flow and the porous substrate at higher $Pr$.

\subsection{Temperature spectrograms}

\begin{figure}
\centerline{\includegraphics{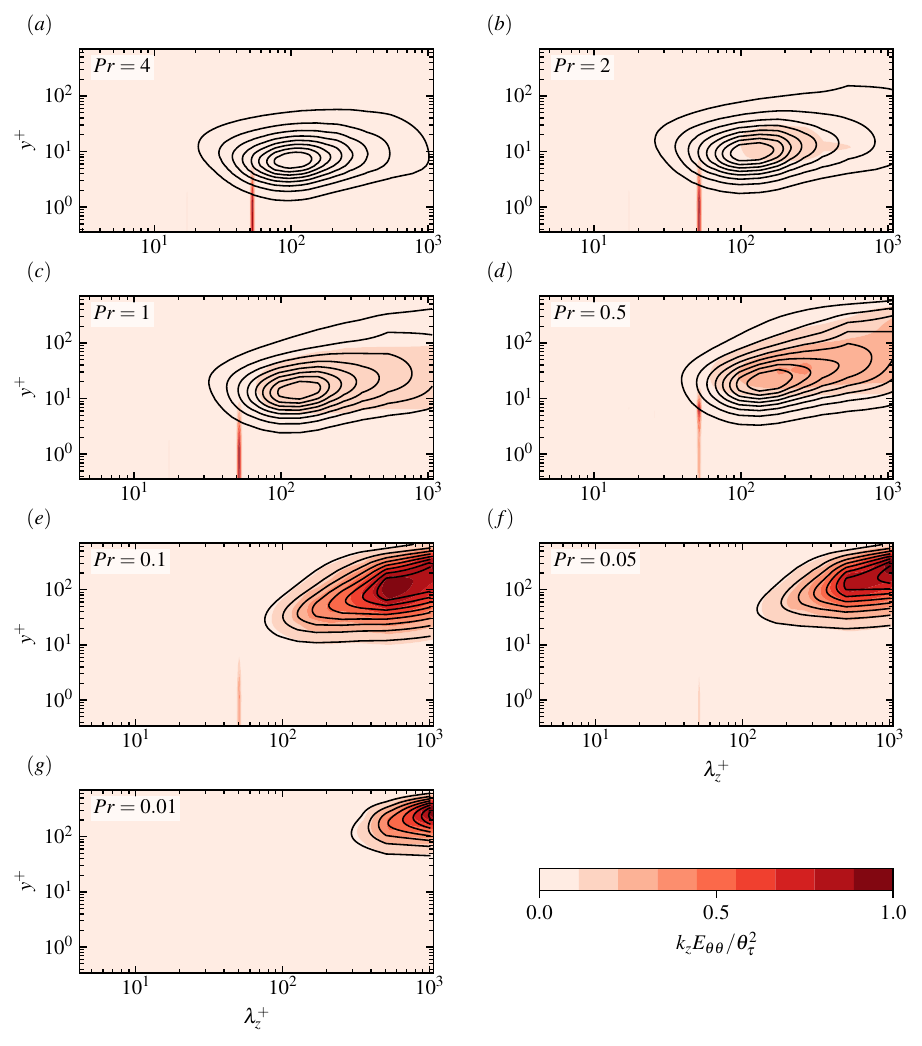}}
  \caption{Premultiplied spectrograms of temperature fluctuations $k_z E_{\theta \theta}/\theta_{\tau}^2$ for cases H1 (\textit{a}), M3 (\textit{b}), M2 (\textit{c}), M1 (\textit{d}), L3 (\textit{e}), L2 (\textit{f}) and L1 (\textit{g}). Spectrograms are normalized by their maximum value, to allow comparison between cases. Black lines show the corresponding smooth-wall spectrograms at matched $Pr$.}
\label{fig:spectra}
\end{figure}

Temperature spectrograms provide insight into how $Pr$ affects the interplay between the large thermal eddies in the core region and the smaller structures in the roughness sublayer. Figure~\ref{fig:spectra} shows the premultiplied spectral densities of temperature, $k_z E_{\theta\theta}/\theta_\tau^2$, as a function of the virtual-origin-adjusted, inner-scaled wall-normal distance $y^+$ and the spanwise wavelength $\lambda_z^+$. Further, the spectrograms have been normalised by the maximum value of $k_z E_{\theta\theta}/\theta_\tau^2$ to enable a comparison of the structure and position of the spectrograms over a wide range of $Pr$.

For the smooth wall at $Pr \geq 1$, the spectrum has a single peak at $\lambda_z^+ \approx 100$ and $y^+ \approx 10$, similar to the behaviour exhibited by the streamwise velocity spectrum in canonical wall-bounded turbulence. This peak corresponds to the spacing of the near-wall streaks and reflects the strong coupling between the thermal and momentum fields at $Pr \sim 1$. As $Pr$ decreases, thermal diffusivity dominates and damps the small-scale thermal fluctuations near the wall; the peak then moves towards the channel centreline, and the thermal fluctuations are carried by large-scale outer structures.
The same trend has been observed in the spectrograms of turbulent pipe flow~\citep{pirozzoli_22} and turbulent channel flow with one-sided heating~\citep{pirozzoli_23}.\par
The porous-wall spectrograms differ most at higher $Pr$, where the peak is confined to a narrow band at $\lambda_z^+ \approx 50$, corresponding to the spanwise pore spacing. We attribute this to a pore-scale, surface-induced mechanism that draws thermal energy from the near-wall broadband turbulence into the porous medium. At high $Pr$, the thermal sublayer is thin, so that the finer near-wall thermal structures can penetrate the porous substrate, shifting the spectrum from broadband to pore-dominated.
With decreasing $Pr$, the pore-scale signature weakens: it is faint from $Pr = 1$ downwards and has almost vanished at $Pr = 0.05$.


The broadband spectra also differ between the porous and smooth walls. The permeable surface shortens the near-wall streaks, so that at $Pr \approx 1$ the porous-wall spectra extend to larger wavelengths and the peak moves away from the wall. In contrast, for $Pr < 0.5$ the smooth and porous-wall spectra largely overlap in shape and location, since the thick conductive sublayer screens the near-wall flow modification and the permeable surface has little influence on the outer flow.

\subsection{Mean temperature statistics}

\begin{figure}
\centerline{\includegraphics{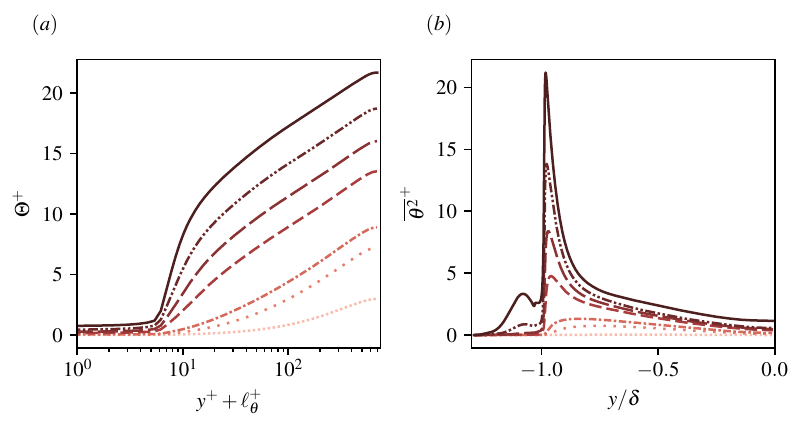}}
  \caption{\textit{(a)} Mean temperature and \textit{(b)} temperature fluctuation profiles for cases $L1$
  (\lsDotted), $L2$ (\lsSparse), $L3$ (\lsDashDot),
$M1$ (\lsDashed),
$M2$ (\lsLong),
$M3$ (\lsDashDotDot) and
$H1$ (\lsSolid).}
\label{fig:meanandfluctemp}
\end{figure}

\begin{figure}
\centerline{\includegraphics{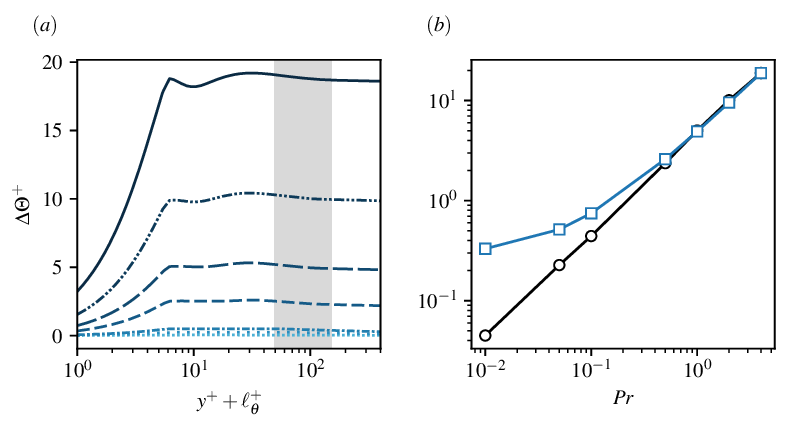}}
  \caption{Temperature deficit $\Delta \Theta^+$ as a function of \textit{(a)} the wall-normal distance for cases $L1$
  (\lsDotted), $L2$ (\lsSparse), $L3$ (\lsDashDot),
$M1$ (\lsDashed),
$M2$ (\lsLong),
$M3$ (\lsDashDotDot) and
$H1$ (\lsSolid), and \textit{(b)} $Pr$. Circles show the DNS data and squares show the prediction of \citet{Chakraborty2026}.}
\label{fig:tempdeficit}
\end{figure}

Figure~\ref{fig:meanandfluctemp}(a) shows the mean temperature profiles for
increasing $Pr$, where two distinct types of profile can be discerned. For
medium-to-high $Pr$ ($Pr > 0.1$), a near-wall roughness sublayer, a
logarithmic layer and a wake region are clearly identifiable. For low
 $Pr$ ($Pr < 0.1$), in contrast, the profile departs from the
canonical logarithmic law and instead takes the form characteristic of a conductive
sublayer. This behaviour is confirmed by figure~\ref{fig:meanandfluctemp}(b),
where the peak of the scalar fluctuations grows in magnitude and moves towards the
wall with increasing $Pr$. At low $Pr$, the high thermal diffusivity smears out the
fine near-wall thermal structures, which are replaced by a conductive sublayer that
extends across nearly half the channel height. As $Pr$ increases, these fine
structures become progressively more energetic and are confined to an increasingly
thin conductive sublayer, which accounts for the growth of the peak fluctuation
intensity. We also see noticeable thermal fluctuations inside the porous medium, for higher $Pr$ ($Pr>0.5$), which also directly supports the phenomenon seen in the instantaneous contours of thermal structures penetrating the porous surface.

Figure~\ref{fig:tempdeficit}(\textit{a}) shows the temperature deficit
$\Delta\Theta^{+} = \Theta^{+}_{\mathrm{S}} - \Theta^{+}_{\mathrm{R}}$
as a function of the wall-normal distance $y^+ +\ell_{\theta}^{+}$. Each curve rises steeply in the
near-wall region and reaches a plateau beyond
$y^+ +\ell_{\theta}^{+}\approx 30$, confirming that the smooth and
porous-wall mean temperature profiles are parallel in the overlap layer and
differ by a constant offset. We evaluate the plateau value by averaging within
the shaded window
($50 \lesssim y^{+}+\ell_{\theta}^{+} \lesssim 150$), with a maximum rms deviation of 0.11 for the highest $Pr$ case. The temperature deficit
increases strongly with $Pr$. At the lowest $Pr$ ($Pr = 0.01$, $0.05$),
the offset almost vanishes and no well-defined plateau exists,
consistent with a conduction-dominated field that lacks a thermal
logarithmic layer, whereas at $Pr = 4$ the plateau lies at
$\Delta\Theta^{+}\approx 18$-$19$. We compare the DNS results for $\Delta\Theta^+$ with the predictive formula of \citet{Chakraborty2026} for the thermodynamic transitionally rough regime,
\begin{align}
   \Delta \Theta^+ = C_1(Pr)\log(k_s^+) + C_2(Pr),
\end{align}
where $C_1 = 2.46Pr+0.53$ and $C_2 = -3.79Pr-1.53$ are fitting constants obtained from DNS data at $Pr=0.5$-$2$. As figure~\ref{fig:tempdeficit}(\textit{b}) shows, the prediction agrees well with the DNS for $Pr>0.1$, well outside the range of Prandtl numbers it was calibrated on, but fails at lower $Pr$. This is because the formula was derived empirically from data at $Pr\sim\mathcal{O}(1)$, where heat transfer is driven by advection, whereas in the low-$Pr$ limit conduction dominates.

\subsection{Sheltered and exposed model for local heat transfer}
We seek to build a predictive model for the roughness heat-transfer coefficient, i.e. the roughness Stanton number, defined as the non-dimensional conductive heat flux in the near-wall region of the porous medium. To this end, we follow \citet{Abu_Rowin_Zhong_Saurav_Jelly_Hutchins_Chung_2024}, who showed that a rough surface can be partitioned into regions exposed to the mean flow and regions sheltered from it.

To quantify this local heat transfer decomposition we can use: 
\begin{align}
    Q_\mathrm{local} &= Q_\mathrm{exposed} + Q_\mathrm{sheltered}, \\
    \frac{\langle{q_w}\rangle}{\rho c_p} &= \underbrace{\frac{1}{A_t}\int_e \alpha \frac{\partial\theta}{\partial n}\bigg|_\mathrm{wall} \mathrm{d}A}_{\text{exposed}} + \underbrace{\frac{1}{A_t}\int_s \alpha \frac{\partial\theta}{\partial n}\bigg|_\mathrm{wall} \mathrm{d}A}_{\text{sheltered}},
\end{align}
where $Q$ is the heat transfer, $\langle {q_w}\rangle$ is the wall heat transfer per unit total plan area, $A_t$ is the plan area of the porous surface, $\theta$ is the local temperature and $n$ is the spatial coordinate normal to the solid surface.
The roughness Stanton number, $C_{h,k}$, then follows as
\begin{align}\label{Chk}
    C_{h,k} &= \frac{\langle{q_w}\rangle}{\rho c_p U_{i} \Theta_{i}} = \frac{1}{U_{i}^+\Theta_{i}^+}
    = \frac{A_e}{A_t}\underbrace{\frac{1}{A_e}\int_e \frac{\alpha}{U_{i} \Theta_{i}} \frac{\partial\theta}{\partial n}\bigg|_\mathrm{wall} \mathrm{d}A}_{C_{h,e}}
    + \frac{A_s}{A_t}\underbrace{\frac{1}{A_s}\int_s \frac{\alpha}{U_{i} \Theta_{i}} \frac{\partial\theta}{\partial n}\bigg|_\mathrm{wall} \mathrm{d}A}_{C_{h,s}} \nonumber \\
    &= \frac{A_e}{A_t}C_{h,e} + \frac{A_s}{A_t}C_{h,s},
\end{align}
where $A_e$ is the exposed wetted area, $A_s$ is the sheltered wetted area, $U_{i}$ is the mean streamwise velocity and  $\Theta_i$ is the mean temperature at the wall-normal location corresponding to the $y/\delta=-1$, at the porous interface. The exposed and sheltered heat transfer contributions are defined as   $C_{h,e}$ and $C_{h,s}$. For sinusoidal roughness, \citet{Abu_Rowin_Zhong_Saurav_Jelly_Hutchins_Chung_2024} modelled the exposed contribution as $C_{h,e}\sim Pr^{-2/3}$, assuming that the conductive sublayer on exposed surfaces behaves similarly to a laminar boundary layer~\citep{Owen_Thomson_1963, Yaglom_Kader_1974}. For the sheltered contribution, they tested both $C_{h,s}\sim Pr^{-0.45}$ and $C_{h,s}=0$, and found the former to be more accurate at larger $k^+$.

\begin{figure}
    \centering
    \includegraphics[width=\linewidth]{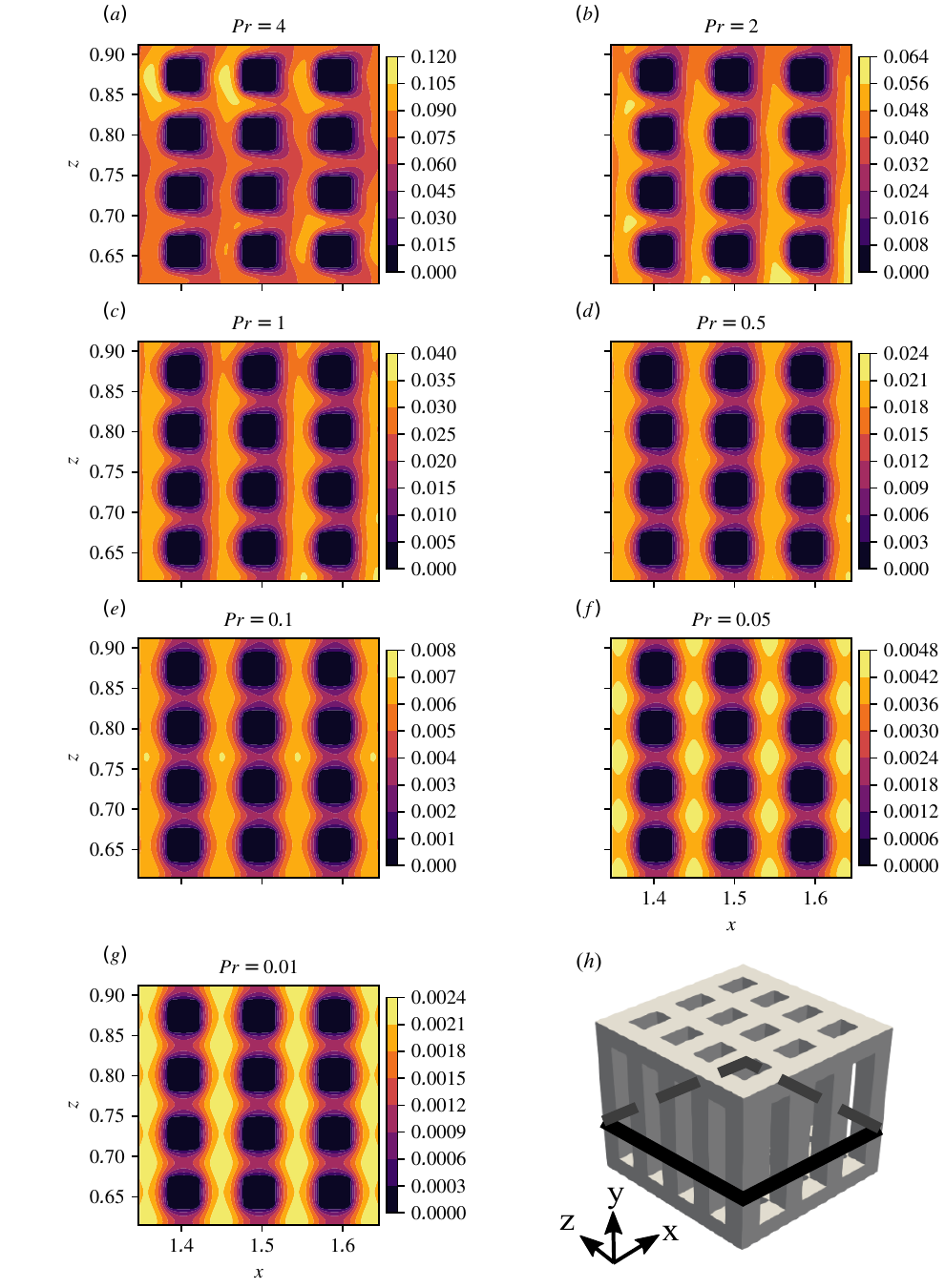}
    \caption{Temporally averaged temperature contours on a wall-parallel plane within the porous layer ($y/\delta=-1.2$), for cases H1 (\textit{a}), M3 (\textit{b}), M2 (\textit{c}), M1 (\textit{d}), L3 (\textit{e}), L2 (\textit{f}) and L1 (\textit{g}). Panel (\textit{h}) shows the $71\%$-porosity unit cell with the slice location at $y/\delta =-1.2$. The flow is along the $x$ direction. }
    \label{fig:theta_slices}
\end{figure}

Figure~\ref{fig:theta_slices} shows contours of the time-averaged temperature
$\theta$ in a wall-parallel ($x$-$z$) plane at the wall-normal location where $U$ reaches a maximum inside the porous surface, at $0.2\delta$ below the surface, 
focusing on a region of $0.3\delta \times 0.3\delta$. The warmer fluid is retained upstream of the constriction between two neighbouring ligaments, whereas a cooler stream bridges the gap directly between the ligaments,
marking the sheltered, recirculation-dominated core of the constriction. These
contours reveal the physical basis of the exposed-sheltered decomposition, and show that it is a meaningful classification only at sufficiently high $Pr$. At $Pr=2$ and
$Pr=4$, the thin thermal sublayer confines strong temperature gradients upstream of each ligament, resulting in different thermal environments on the windward and leeward sides. The exposed lateral surfaces, swept by fluid
entering the passages, carry steep near-wall gradients and the high temperature regions,
whereas the sheltered surfaces bounding the recirculating throat lie in the cooler sheltered region, where gradients are much gentler. The decomposition quantifies this difference in the thermal field. As $Pr$ decreases, diffusion smears out these gradients. The isotherms thicken and
relax away from the walls; the warmer regions broaden and merge across the throat. At $Pr=0.05$ and $0.01$, the
field is smooth, symmetric and nearly uniform across the constrictions. The exposed and
sheltered surfaces experience similar thermal environments and transfer heat by the
same conduction-limited mechanism; the geometric distinction between the two faces no longer produces distinctly different thermal fields. This transition 
reflects the change from advection-dominated to
diffusion-dominated thermal transport with lowering $Pr$. It is also consistent with the convergence of
$C_{h,e}$ and $C_{h,s}$ onto a common $Pr^{-1}$ scaling as $Pr\to0$, as discussed
below.

We apply the same methodology to the cubic-lattice porous surfaces. Each wetted surface element is classified using the time-averaged streamwise velocity $\overline{u_t}$ at the first fluid node adjacent to it: elements with $\overline{u}_t>0$ are classified as exposed and the remaining as sheltered. On faces aligned with the flow, this distinguishes forward from reversed flow, as in \citet{Abu_Rowin_Zhong_Saurav_Jelly_Hutchins_Chung_2024}; on faces normal to the streamwise direction, it separates windward (impinging) from leeward faces. We consider the lattice with $\Phi = 71\%$ at $Pr = 0.01$, $0.05$, $0.1$, $0.5$, $1$, $2$ and $4$, and the lattice with $\Phi = 87\%$ at $Pr = 0.5$, $1$ and $2$. Relative to the wetted area $A_w$, i.e.\ the total surface area of the porous substrate, the exposed and sheltered fractions are $A_e/A_w = 0.68$ and $A_s/A_w = 0.32$ for $\Phi = 71\%$, and $A_e/A_w = 0.78$ and $A_s/A_w = 0.22$ for $\Phi = 87\%$. The total plan area is $A_t = L_x \times L_z = 4.5\delta^2$, where $L_x$ and $L_z$ are the streamwise and spanwise extents of the porous surface.

\begin{figure}
    \centering
    \includegraphics{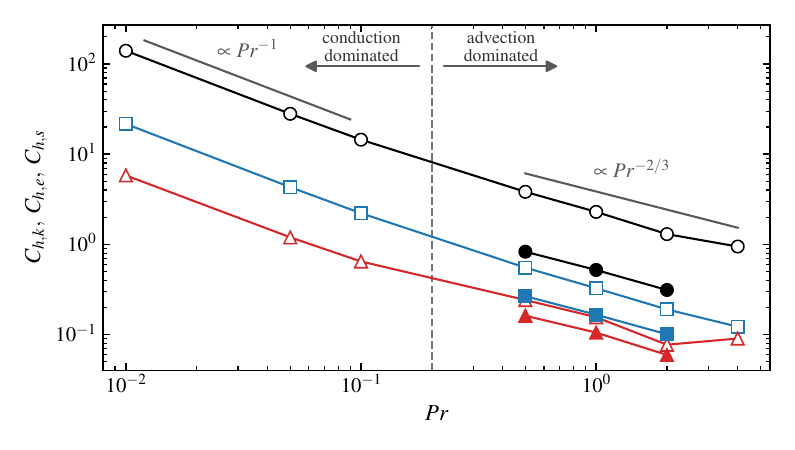}
    \caption{Local roughness heat-transfer coefficients as a function of $Pr$: total $C_\mathrm{h,k}$ (black markers), exposed $C_\mathrm{h,e}$ (blue markers) and sheltered $C_\mathrm{h,s}$ (red markers). Lines with filled markers show cases from $87\%$ porosity and lines with white markers show cases from $71\%$ porosity.}
    \label{fig:localStantonnumber}
\end{figure}

For $\Phi = 71\%$ and $Pr<0.1$, all three local heat-transfer coefficients, $C_{h,k}$, $C_{h,e}$ and $C_{h,s}$, scale as $Pr^{-1}$ (figure~\ref{fig:localStantonnumber}). This scaling reflects the transition from advection to conduction-dominated transport near the solid surfaces. As $Pr$ decreases, the conductive sublayer becomes thicker until it exceeds the pore and ligament scales. The conduction in the domain is then controlled by a geometry-defined length scale. Within this layer the mean temperature follows $\Theta^+ = Pr\,y^+$, and from equation~\eqref{Chk}, $C_{h,k} = 1/(U_i^+\Theta_i^+)$. Since $U_i^+$ is independent of $Pr$, $C_{h,k}\sim 1/\Theta_i^+ \sim Pr^{-1}$, and the same argument applies to $C_{h,e}$ and $C_{h,s}$, as both exposed and sheltered surfaces lie within the same conductive layer.  However, the bulk Nusselt number has not yet saturated at the lowest $Pr$ considered (table~\ref{tab:results}), since turbulent mixing in the channel core remains active at $Pe_b = Re_b\,Pr = \mathcal{O}(10^2)$. The $Pr^{-1}$ scaling of the local coefficients therefore marks the onset of conduction-limited heat transfer at the porous surface, which precedes the fully conductive limit of the bulk flow.\par

For $Pr>0.1$, $C_{h,e}\sim Pr^{-0.73}$ and $C_{h,s}\sim Pr^{-0.52}$, the latter being close to the $Pr^{-0.45}$ scaling reported for rough walls. The overall coefficient follows $C_{h,k}\sim Pr^{-0.68}$, consistent with rough-wall behaviour~\citep{Abu_Rowin_Zhong_Saurav_Jelly_Hutchins_Chung_2024}. In this medium to high $Pr$ range, the $Pr^{-2/3}$ scaling originates from the Reynolds analogy.\par

At the higher porosity, $\Phi = 87\%$, the exposed and sheltered coefficients, and therefore $C_{h,k}$, all scale as $Pr^{-0.7}$ over $Pr=0.5$-$2$, close to the $Pr^{-2/3}$ dependence. Although they describe hydrodynamically distinct portions of the wetted surface, the exposed and sheltered coefficients share a common exponent, indicating that both are controlled by the same near-wall conductive-sublayer mechanism. The most notable feature at this porosity is the sheltered scaling, which shows a markedly stronger dependence on $Pr$ than the $Pr^{-0.52}$ and $Pr^{-0.45}$ scalings found for the $71\%$ lattice and for rough walls, respectively. We attribute this to the larger pores of the $87\%$ lattice, whose open structure facilitates higher suction and blowing and disrupts the sheltering on the ligaments. The sheltered regions are therefore no longer stagnant and conduction-dominated, but behave thermally much like the exposed surfaces, so that all three coefficients collapse onto a single $Pr^{-2/3}$ law.
\subsubsection{Propagating to bulk Stanton number from $C_{h,k}$}
We now propagate the local coefficient $C_{h,k}$ to the bulk Stanton
number. In the fully rough regime, the mean velocity in the logarithmic region
follows Nikuradse's fully rough asymptote,
\begin{equation}\label{eq:nik}
  U^+ = \frac{1}{\kappa}\log\left(\frac{y}{k_s}\right) + C_N ,
\end{equation}
where $y$ here denotes the distance from the wall, $\kappa = 0.387$ is
the von K\'arm\'an constant and $C_N \approx 8.5$
is the fully rough intercept. The mean temperature follows an analogous
logarithmic law,
\begin{equation}\label{eq:thetalog}
  \Theta^+ = \frac{1}{\kappa_\theta}\log\left(\frac{y}{k_s}\right) + g(k_s^+,Pr),
\end{equation}
where $\kappa_\theta = 0.459$ and $g$ is the heat-transfer roughness
function, which can be interpreted as the log-law temperature at
$y = k_s$, $g = \Theta^+_{k_s}$
\citep{Dipprey1963, Owen_Thomson_1963, Zhong_Hutchins_Chung_2023,
Abu_Rowin_Zhong_Saurav_Jelly_Hutchins_Chung_2024}.

We define the friction coefficient and Stanton number  based on the centreline
quantities,
\begin{equation}\label{eq:defs_delta}
  C_{f,\delta} = \frac{2\tau_w}{\rho U_\delta^2}, \qquad
  St_\delta = \frac{q_w}{\rho c_p U_\delta \Theta_\delta},
\end{equation}
where $U_\delta$ and $\Theta_\delta$ are the log-law velocity and
temperature evaluated at the channel centreline. Further,
$U_\delta^+ = \sqrt{2/C_{f,\delta}}$ and
$St_\delta = 1/(U_\delta^+\Theta_\delta^+)$. Evaluating equations~\eqref{eq:nik} and
\eqref{eq:thetalog} at the centreline gives
\begin{align}
  \sqrt{2/C_{f,\delta}} &= U_\delta^+
  = \frac{1}{\kappa}\log\left(\frac{\delta}{k_s}\right) + C_N , \label{eq:uh}\\
  \frac{\sqrt{C_{f,\delta}/2}}{St_\delta} &= \Theta_\delta^+
  = \frac{1}{\kappa_\theta}\log\left(\frac{\delta}{k_s}\right) + g . \label{eq:thetah}
\end{align}
Eliminating $\log(\delta/k_s)$ by combining equations~\eqref{eq:uh} and \eqref{eq:thetah} yields
\begin{equation}\label{eq:combined}
  \kappa\left(\sqrt{2/C_{f,\delta}} - C_N\right)
  = \kappa_\theta\left(\frac{\sqrt{C_{f,\delta}/2}}{St_\delta} - g\right),
\end{equation}
which can be rearranged as
\begin{equation}\label{eq:Sth}
  St_\delta = \frac{C_{f,\delta}/2}
  {(\kappa/\kappa_\theta)\left(1 - C_N\sqrt{C_{f,\delta}/2}\right)
   + g\sqrt{C_{f,\delta}/2}} .
\end{equation}
Equation~\eqref{eq:Sth} is a generalized Reynolds analogy: for a given
$C_{f,\delta}$, the Stanton number depends on the roughness only through
$g$, with no explicit dependence on $\delta/k_s$. If the momentum and thermal
log laws were identical ($\kappa_\theta = \kappa$ and $g = C_N$), equation~\eqref{eq:Sth} would reduce to the classical Reynolds analogy,
$St_\delta = C_{f,\delta}/2$.

We use $C_{h,k}$ as defined in equation~\eqref{Chk}, which is based on the
interface values $U_i^+$ and $\Theta_i^+$. The log laws \eqref{eq:nik}
and \eqref{eq:thetalog} give $U^+_{k_s} = C_N$ and $\Theta^+_{k_s} = g$, so $g$ is related to $C_{h,k}$ by
\begin{equation}\label{eq:g_Chk}
  g = \frac{1}{C_N\, r_U\, r_T\, C_{h,k}}, \qquad
  r_U = \frac{U_i^+}{C_N}, \quad r_T = \frac{\Theta_i^+}{g},
\end{equation}
where $r_U$ and $r_T$ measure the ratio of  interface values  to 
the log-law values at $y = k_s$ for velocity and temperature. Substituting \eqref{eq:g_Chk} into
\eqref{eq:Sth} expresses the $St_{\delta}$ directly in terms
of $C_{h,k}$,
\begin{equation}\label{eq:St_delta_Chk}
  St_\delta = \frac{C_{f,\delta}/2}
  {(\kappa/\kappa_\theta)\left(1 - C_N\sqrt{C_{f,\delta}/2}\right)
   + \sqrt{C_{f,\delta}/2}\,\big/\big(C_N\, r_U\, r_T\, C_{h,k}\big)} .
\end{equation}
To express the result in terms of bulk quantities, we define
\begin{equation}\label{eq:defs_bulk}
  C_{f,b} = \frac{2\tau_w}{\rho U_b^2}, \qquad
  St_b = \frac{q_w}{\rho c_p U_b \Theta_m},
\end{equation}
where $U_b$ is the bulk velocity and
$\Theta_m = \int_{-\delta}^{\delta} U\Theta\,\mathrm{d}y \big/ \int_{-\delta}^{\delta} U\,\mathrm{d}y$
is the mixed-mean temperature. Because both sets of definitions use the
same $\tau_w$ and $q_w$, they are related exactly by
\begin{equation}\label{eq:convert}
  C_{f,b} = C_{f,\delta}\left(\frac{U_\delta^+}{U_b^+}\right)^2, \qquad
  St_b = St_\delta\,\frac{U_\delta^+\,\Theta_\delta^+}{U_b^+\,\Theta_m^+} .
\end{equation}

\begin{figure}
    \centering
    \includegraphics[width=\linewidth]{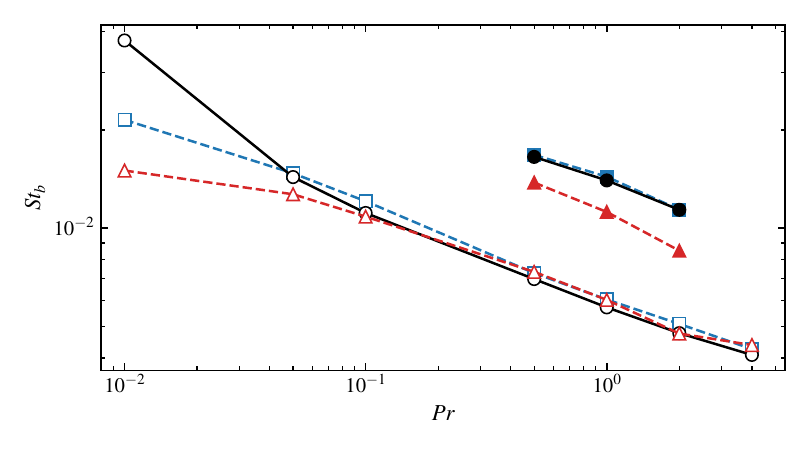}
    \caption{Bulk Stanton number $St_b$ from DNS (black lines), from the exposed-sheltered predictive model (red line) and from the predictive model of \citet{Chakraborty2026} (blue line). Lines with filled markers show cases from $87\%$ porosity and lines with white markers show cases from $71\%$ porosity.}
    \label{fig:St_prediction}
\end{figure}

Propagating the local coefficients to the bulk Stanton number through the relation of \citet{Abu_Rowin_Zhong_Saurav_Jelly_Hutchins_Chung_2024} reproduces the DNS data well for the $71\%$ lattice across almost the entire $Pr$ range, the only exception being $Pr=0.01$, where the prediction underestimates $St_b$, see figure~\ref{fig:St_prediction}. This relation is based on the Reynolds analogy and links the bulk transport to a turbulent thermal logarithmic layer through $C_{h,k,\mathrm{log}}=1/(C_N g)$ and the velocity logarithmic law. At $Pr=0.01$, the flow is dominated by conduction and no thermal logarithmic region exists: the mean temperature profile is set by molecular conduction rather than by a canonical turbulent boundary-layer profile, so that the fitted thermal intercept $g$ loses its physical meaning and the prediction cannot capture the conduction-limited enhancement of $St_b$. At all higher $Pr$, where a thermal logarithmic layer is present, the prediction closely follows the DNS data.\par

\begin{figure}
    \centering
    \includegraphics[width=\linewidth]{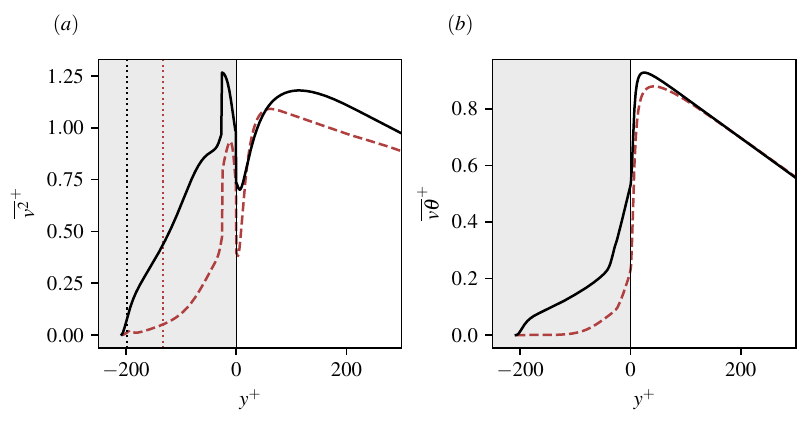}
    \caption{\textit{(a)} Wall-normal fluctuation and \textit{(b)} wall-normal turbulent heat flux for case $M2$ (red-dashed line) and $M2-87$ (black solid line). The dotted black line shows the penetration depth for $M2-87$ (position where the $\overline{v^2}^+$ falls to $10\%$ of the interface value) and the dotted red line for $M2$ in \textit{(a)}. The shaded grey area shows the substrate region.   }
    \label{fig:vfluc_vtheta}
\end{figure}

For the $87\%$ porous lattice, $St_b$ is systematically underpredicted across the full range $0.5 \le Pr \le 2$ (figure~\ref{fig:St_prediction}). The deviation persists across all $Pr$, and we can attribute it clearly to the high permeability of the surface. At $87\%$ porosity, the overlying turbulence penetrates the large pores and advects heat through the open structure, reaching ligament faces that would otherwise be sheltered. This is supported by figure~\ref{fig:vfluc_vtheta}\textit{(a)}: the wall-normal velocity fluctuations $\overline{v^2}^+$ peak higher at the fluid-porous interface and also penetrate deeper into the substrate for $87\%$ than for $71\%$ porosity. These fluctuations carry heat into the substrate, as shown by the correspondingly larger and deeper-reaching turbulent heat flux $\overline{v\theta}^+$ in figure~\ref{fig:vfluc_vtheta}\textit{(b)}. This additional pore-scale advection, the same mechanism that renders the nominally sheltered regions thermally exposed, is not captured by equation~\eqref{eq:Sth}. As a result, the predicted $St$ is systematically too low, and we expect the deviation to grow with increasing permeability. 

\section{Conclusion}

We have studied local heat transfer within porous substrates using direct numerical
simulations of turbulent channel flow at
$Re_\tau \approx 700$ over two cubic-lattice substrates, spanning $Pr = 0.01$-$4$ at $\Phi = 71\%$ and
$Pr = 0.5$-$2$ at $\Phi = 87\%$. Following
\citet{Abu_Rowin_Zhong_Saurav_Jelly_Hutchins_Chung_2024}, we decomposed the local
roughness heat-transfer coefficient $C_{h,k}$ into contributions from surfaces exposed
to and sheltered from the mean flow, and propagated the result to the bulk Stanton
number $St$. Two main conclusions emerge.

First, the rough-wall scaling laws carry over to porous lattices of moderate porosity,
provided that heat transfer is advection dominated. For $\Phi = 71\%$ and $Pr > 0.1$, the
exposed coefficient scales as $C_{h,e} \sim Pr^{-0.73}$ and the sheltered coefficient
as $C_{h,s} \sim Pr^{-0.52}$, close to the $Pr^{-2/3}$ Reynolds-analogy and
$Pr^{-0.45}$ laws repo      v   rted for rough walls, whereas the overall heat transfer coefficient follows
$C_{h,k} \sim Pr^{-0.68}$. At the higher porosity, $\Phi = 87\%$, the exposed surfaces
still behave as on a rough wall; however, the sheltered contribution departs from
rough-wall behaviour, following $C_{h,s} \sim Pr^{-0.7}$ instead of $Pr^{-0.45}$. As a
result, all three coefficients collapse onto a single law close to $Pr^{-2/3}$. The
larger pores allow the flow to advect heat onto the leeward faces of the ligaments, so
that the sheltered regions are no longer stagnant and conduction-limited, but behave
thermally like exposed surfaces. At low $Pr$ ($Pr < 0.1$), a different
scaling emerges altogether, with $C_{h,k}$, $C_{h,e}$ and $C_{h,s}$ all scaling as
$Pr^{-1}$. In this regime, heat transfer at the porous surface is conduction-dominated: the conductive sublayer exceeds the pore scale, the conduction length is set by the lattice geometry, and the local coefficients scale as $Pr^{-1}$, even though the bulk Nusselt number has not yet saturated. The temperature field within the pores becomes smooth and nearly uniform, so that the distinction between exposed and sheltered surfaces no longer has a thermal meaning.

Second, the bulk Stanton number can be predicted from the local coefficient $C_{h,k}$.
For $\Phi = 71\%$, propagating $C_{h,k}$ through the rough-wall relation of
\citet{Abu_Rowin_Zhong_Saurav_Jelly_Hutchins_Chung_2024} reproduces the DNS data
closely over almost the entire $Pr$ range. The exception is $Pr = 0.01$, for
which no thermal logarithmic layer exists and the fitted intercept $g$ loses its
physical meaning. For $\Phi = 87\%$, however, the exposed-sheltered prediction
consistently underestimates $St_b$ across $Pr = 0.5$-$2$, because the relation does
not account for the pore-scale transport that makes the
sheltered faces thermally active. In
contrast, the predictive model of \citet{Chakraborty2026} matches the DNS data at both
porosities throughout the advection-dominated range. This model therefore remains the
more robust tool for predicting bulk heat transfer over highly permeable porous
surfaces, whereas the exposed-sheltered decomposition is useful for explaining
the local physics underlying it.

\textbf{Acknowledgements.} The work by A.C. is supported by TU Delft AI labs and Talent Programme. A.C. acknowledges SURF (www.surf.nl) for the support in using the Dutch National Supercomputer Snellius through grants EINF-17240 and EINF-12445. D.M. acknowledges funding from the Italian Ministry of University and Research (MUR) through the FIS2 project HEATFORCE (grant no. FIS-2023-00346; CUP D53C25002050001).

\textbf{Use of artificial intelligence.} The authors used Claude (Anthropic, model Claude Opus 5.5) between September and November 2026 to draft and revise parts of the text, and to write Python scripts that produce figures from simulation datasets. All generated text and plots were verified by the authors. No figure was modified using AI beyond the plotting of data.




\bibliography{sn-bibliography}

\end{document}